\documentstyle[preprint,aps]{revtex}
\begin{document}
\draft
\title{NON--MINIMAL SCALAR--TENSOR THEORIES AND QUANTUM GRAVITY}
\author{FATIMAH SHOJAI\thanks{Email: FATIMAH@THEORY.IPM.AC.IR}}
\address{Department of Physics, Iran University of Science and Technology, P.O. Box 16765--163,
Narmak, Tehran, IRAN,}
\address{\&}
\address{Institute for Studies in Theoretical Physics and Mathematics,
P.O. Box 19395--5531, Tehran, IRAN.}
\author{Ali SHOJAI\thanks{Email: SHOJAI@THEORY.IPM.AC.IR}}
\address{Department of Physics, Tarbiat Modares University, P.O. Box 14155--4838,
Tehran, IRAN,}
\address{\&}
\address{Institute for Studies in Theoretical Physics and Mathematics,
P.O. Box 19395--5531, Tehran, IRAN.}
\date{22 September 1999}
\maketitle
\newpage
\begin{abstract}
\begin{center}
\parbox{15cm}{\it Recentely\cite{GEO} it is shown that the quantum effects of matter determine the conformal degree of freedom of the space--time metric. This was done in the framework of a scalar--tensor theory with
one scalar field\cite{CON,STQ}. A point with that theory is that the form of quantum potential is preassumed.
Here we present a scalar--tensor theory with two scalar fields, and no assumption on the form of quantum potential. It is shown that using the equations of motion one gets the correct form of quantum potential plus some corrections.}
\end{center}
\end{abstract}
\vspace{1cm}
\pacs{04.60.--m, 98.80.Hw, 03.65.BZ}
\section{INTRODUCTION}
Scalar--tensor theories are used frequently in the investigation of gravity. Among these theories a
theory with two scalar fields is more useful. Two examples are superstring and Kaluza--Klein theories.
Superstring theory is a good candidate for quantum gravity, that is more fruitful
 in ten dimension. For bringing the theory in four dimensions, the extra six dimensions must be compactified. At the low energy limit, the gravitational interaction emerges from the theory. In addition to the metric field, this gravitational interaction includes two other fields.
The low energy effective action in four dimension is given by:
\begin{equation}
{\cal A}=\int \! d^4x\ \sqrt{-g}\left [ \phi{\cal R}-\omega\frac{\nabla_\mu\phi\nabla^\mu\phi}{\phi} -\frac{\nabla_\mu\chi\nabla^\mu\chi}{\phi}\right ]
\label{a1}
\end{equation}
In this action one scalar field is coupled non--minimally to gravity as in the Brans--Dicke theory. The
second scalar field is coupled minimally to gravity, but has a non--trivial coupling with the first scalar
field.  Note that $\omega$ is the coupling constant which is equal to $-1$ for the string theory.
The frame in which this action is written is called the string frame (or Jordan frame).
The above  action can be rewritten as Einstein--Hilbert action, by the following conformal transformation:
\begin{equation}
\widetilde{g}_{\mu\nu}=\phi^{-1}g_{\mu\nu}
\label{a2}
\end{equation}
In the transformed frame (the Einstein frame), the action functional is given by:
\begin{equation}
{\cal A}=\int \! d^4x\ \sqrt{-\widetilde{g}}\left [ \widetilde{{\cal R}} -(\omega+3/2)\frac{\widetilde{\nabla}_\mu\phi\widetilde{\nabla}^\mu\phi}{\phi^2}
-\frac{\widetilde{\nabla}_\mu\chi\widetilde{\nabla}^\mu\chi}{\phi^2} \right ]
\end{equation}
Note that, in this action, both the scalar fields are coupled minimally to gravity. Although the Einstein frame is frequently used, but the question that the physical interpretation must be presented in which  frame, is an open problem\cite{FAR,LEV}.

Furthur investigation of the low energy string theory action, its particular solutions, and its symmetries can be found in \cite{BAR}.

In addition, in the Kaluza--Klein theory, the effective action is the same as above. The bosonic sector
in the Kaluza--Klein supergravity,  after compactification , includes metric and two scalar fields. In fact the action would be just like the equation (\ref{a1}). But the coupling constant $\omega$ is now equal to $\frac{1-d}{d}$ where $n=4+d$ is the space--time dimension.

Since both the string theory and Kaluza--Klein theory after going to the low energy regime and reduction of dimension, leads to a non--minimal coupling,
quantization of non--minimal theory with action (\ref{a1}), is very important.

In references \cite{FAB,COL} the cosmological aspects of this model are investigated. For this purpose, two
effective models are considered in \cite{COL}. As a first model a minimally coupled scalar field in a FRW universe is chosen. The other model consists of two minimally coupled scalar fields related to the string theory or Kaluza--Klein theory through a conformal transformation. Both these models are
quantized.
In another work\cite{FAB}, a non--minimall coupling is considered. As the Wheeler--DeWitt equation (WDW equation) is not separable for the non--minimal coupling, the conformal transformation given by equation (\ref{a2}) yeilds a separable WDW equation. Thus even the quantum interpretation requires the conformal transformation and the quantum behaviour of the physical quantities is obtained in the Einstein frame. The result should be transformed back to the original frame.

In both the above works causal interpretation of quantum theory (de-Broglie--Bohm theory) is adopted. Bohmian trajectories are investigated and it is shown that for some exact solutions, the classical limit appears for small scale factor of the universe. Therefore the classical initial singularity is not removed.

Recently \cite{GEO} we have shown that there is  a very close connection between the causal version
of quantum mechanics and the geometry of the space--time. In fact, the quantum potential which
would be explained in the next section determines the conformal factor of the space--time metric.
It is shown that this idea can be realized in the framework of a scalar--tensor theory with one
scalar field\cite{CON,STQ}. A point with that theory is that the form of quantum potential is preassumed.
In this work we shall show that using two scalar fields one can relax this preassumption and
on the equations of motion the correct form of quantum potential will be achieved.
\section{QUANTUM MECHANICS AND GEOMETRY}
We use causal interpretation of quantum mechanics\cite{BOH,HOL} (de-Broglie--Bohm theory) for quantum gravity, because of its advantages, specially in quantum gravity and cosmology. This theory represents a realistic and deterministic picture of physical phenomena. All particles have a definite trajectory which can be evaluated by the initial conditions. The probabilistic results of this theory are consistent with the standard (Copenhagen) quantum mechanics. A new character in de-Broglie--Bohm theory is the {\it quantum potential\/} distinguishing between classical and quantum regimes.  As in the classical regime one uses the classical Hamilton--Jacobi equation, in the quantum regime one should set:
\begin{equation}
\frac{\partial S}{\partial t}+\frac{|\vec{\nabla}S|^2}{2m} +V+Q=0
\label{a3}
\end{equation}
where $S$ is the Hamilton function, $V$ is the classical potential, and $Q$ is the quantum potential which is a function of ensemble density ($\rho$):
\begin{equation}
Q=-\frac{\hbar^2}{2m}\frac{\nabla^2\sqrt{\rho}}{\sqrt{\rho}}
\end{equation}
In an equivalant manner, one can write down the second law of Newton including the quantum potential:
\begin{equation}
m\frac{d^2\vec{x}}{dt^2}=-\vec{\nabla}(V+Q)
\end{equation}
Another equation of this theory is the continuity equation:
\begin{equation}
\frac{\partial\rho}{\partial t}+\vec{\nabla}\cdot\left (\frac{\vec{\nabla}S}{m}\rho\right )=0
\label{a4}
\end{equation}
It must be noted that a canonical transformation of the form:
\begin{equation}
\psi=\sqrt{\rho}\exp [iS/\hbar]
\label{a5}
\end{equation}
converts the equations (\ref{a3}) and (\ref{a4}) to the Schr\"odinger equation.

Quantum potential includes information about environment and boundaries. In addition, it has a nonlocal nature, any change in the environment or in other particles (in the case of many particle systems\cite{BOH,HOL}) affects the particle. The value of quantum potential and its gradiant define the classical limit. For this limit it is necessary that:
\begin{equation}
Q\ll V\ \ \ \ \ \ \ \ ;{\rm and} \ \ \ \ \ \ \ \ \vec{\nabla}Q\ll \vec{\nabla}V
\end{equation}

The extension of de-Broglie--Bohm theory to fields is straightforward\cite{HOL}. For example in Bohmian quantum gravity, the metric field is the dynamical degree of freedom, and one deals with two equations of motion, the Einstein--Hamilton--Jacobi equation and the conservation equation for probability in the superspace. Many physical results can be deduced from this approach to quantum gravity and cosmology. Interpretation of universe as an individual system, investigation of the general covariance at the quantum level, the emergance of time parameter, and so on, are some of the important results of Bohmian quantum gravity\cite{HOL,HOR,SHT,QGC}.

Solutions to Bohmian quantum gravity equations can be found more easily in some minisuperspace. The time evolution of the metric for many minisuperspaces and investigation of problems like the classical limit, singularity, can be obtained in the literature\cite{HOR}.

Recently\cite{GEO}, we combine the gravity and de-Broglie--Bohm quantum theory in a different manner. It is shown that the matter quantum effects can be included in the space--time geometry and thus they have geometrical nature.  In order to see how this is possible, we must first write down the relativistic Bohmian equations of motion.
Extension of the Hamilton--Jacobi equation is straightforward:
\begin{equation}
\partial_\mu S\partial ^\mu S=m^2c^2(1+Q)={\cal M}^2c^2
\label{a6}
\end{equation}
with
\begin{equation}
Q=\alpha\frac{\Box\sqrt{\rho}}{\sqrt{\rho}};\ \ \ \ \ \ \ \ \alpha=\frac{\hbar^2}{m^2c^2}
\label{a21}
\end{equation}
and the continuity equation is now:
\begin{equation}
\partial_\mu(\rho\partial^\mu S)=0
\end{equation}
Again a canonical transformation of the form (\ref{a5}), leads to the Klein--Gordon equation.
The  mass quantity ${\cal M}$ modified by quantum potential with respect to the classical mass $m$, is called the quantum mass. The Hamilton--Jacobi equation (\ref{a6}) would be transformed to the classical one, if the space--time metric is changed by the conformal transformation from $\overline{g}_{\mu\nu}$ to:
\begin{equation}
g_{\mu\nu}=\frac{{\cal M}^2}{m^2}\overline{g}_{\mu\nu}
\end{equation}
In this way the matter quantum effects are included in the conformal factor of the space--time metric. From this view point the $g_{\mu\nu}$ metric is the physical metric, which has an extra space--time dependence through the quantum potential. In reference \cite{GEO}, the back reaction of the conformal factor on the background metric is considered to complete the theory.  In this work we have started with the Einstein--Hilbert action functional plus the classical matter action (no quantum potential is considered), and then the above mentioned conformal transformation is used to introduce the quantal behaviour of matter. The quantum gravity equations are obtained from the transformed action. They are:
\begin{equation}
\overline{\nabla}_{\mu}\left (\rho \Omega^2 \overline{\nabla}^{\mu}S \right )=0
\end{equation}
\begin{equation}
\overline{\nabla}_{\mu}S \overline{\nabla}^{\mu}S =m^2\Omega^2
\end{equation}
\[ \Omega^2\left [ \overline{{\cal R}}_{\mu \nu}-\frac{1}{2}\overline{g}_{\mu \nu}\overline{{\cal R}}\right ]
-\left [ \overline{g}_{\mu \nu}\stackrel{-}{\Box} -\overline{\nabla}_{\mu}\overline{\nabla}_{\nu}\right ]
\Omega^2 -6 \overline{\nabla}_{\mu}\Omega \overline{\nabla}_{\nu}\Omega+3\overline{g}_{\mu \nu}\overline{\nabla}_{\alpha}\Omega \overline{\nabla}^{\alpha}\Omega\]
\begin{equation}
+\frac{2\kappa}{m}\rho\Omega^2 \overline{\nabla}_{\mu}S
\overline{\nabla}_{\nu}S-\frac{\kappa}{m}\rho \Omega^2\overline{g}_{\mu
\nu} \overline{\nabla}_{\alpha}S \overline{\nabla}^{\alpha}S +\kappa m \rho \Omega^4 \overline{g}_{\mu \nu}=0
\end{equation}
\begin{equation}
\Omega^2=1+\alpha\frac{\stackrel{-}{\Box}\sqrt{\rho}}{\sqrt{\rho}}
\end{equation}
In the above equations, $\overline{g}_{\mu\nu}$ is the background metric, and $\Omega^2$ is the conformal factor of the space--time metric.
It must be noted that in the action leading to the above equations, the dependence of conformal factor on the quantum potential is inserted by hand, using the method of lagrange multipliers.
Also, the dependence of quantum potential on the ensemble density is preassumed.
The above equations are then applied to cosmology. For a radiation--dominated Friedmann--Robertson--Walker background metric, the solution can be derived analyticaly and the initial singularity of the universe can be removed by the quantum effects. Also the correct classical limit is given. In addition to the advantages of Bohiam quantum gravity mentioned previously, it must be noted that this is the de-Broglie--Bohm theory which enables us to get the above theory of quantum gravity. Although in this approach these are the quantal effects of matter which are considered, but the theory is in fact a quantum gravity theory from a Machian viewpoint. What is called the geometry of the space--time, contains both the gravitational and the quantal effects of the matter. Furthuremore, the nonlocal aspect of the quantum potential causes any change in the matter distribution to the geometry of the space--time be influenced simultaneousely\cite{NOL}.
It must be noted that the physical meaning of the conformal factor of the
 space--time metric is given explicitely in this theory.

At the next step, for completing this new approach to quantum gravity we have
used
the scalar-tensor theories\cite{STQ}. This is in order to write an appropriate
 action such
that the conformal factor could be assumed as a dynamical field. Thus
 it is necessary that the equation of motion of the scalar field be in
agreement with the relation between the conformal factor and quantum
potential exactly
(or at least at the first order of approximation). Therefore
 as it is shown in reference\cite{STQ},
there is no need to use the lagrange multiplier.

The matter lagrangian represents an ensemble of relativistic particles.
Each term of it is coupled with the scalar field through an arbitrary
power of it, for
simplicity. These powers are fixed by physical reasons finally.
Also Machian intuition (the relation between global and local structure
of the univrese) leads us to consider some interaction
 between cosmological constant and quantum potential.

With above explanations, we have used the appropriate action as:
\begin{equation}
{\cal A}=\int d^4x \sqrt{-g}\left \{ \phi {\cal R}-\frac{\omega}{\phi}\nabla^\mu\phi
\nabla_\mu\phi+2\Lambda\phi+{\cal L}_m\right \}
\end{equation}
with:
\begin{equation}
{\cal L}_m=\frac{\rho}{m}\nabla^\mu S\nabla_\mu S-m\rho\phi-\Lambda(1+Q)^2
\end{equation}
that leads to the equations of motion:
\begin{equation}
\phi=1+Q-\frac{\alpha}{2}\Box Q
\end{equation}
\begin{equation}
\nabla^\mu S\nabla_\mu S=m^2\phi-\frac{2\Lambda m}{\rho}(1+Q)(Q-\tilde{Q})
+\frac{\alpha\Lambda m}{\rho}\left ( \Box Q -2\nabla_\mu Q\frac{\nabla^\mu
\sqrt{\rho}}{\sqrt{\rho}}\right )
\label{a7}
\end{equation}
\begin{equation}
\nabla^\mu(\rho\nabla_\mu S)=0
\end{equation}
\begin{equation}
{\cal G}^{\mu\nu}-\Lambda g^{\mu\nu}=-\frac{1}{\phi}{\cal T}^{\mu\nu}
-\frac{1}{\phi}[\nabla^\mu\nabla^\nu-g^{\mu\nu}\Box ]\phi+\frac{\omega}{\phi^2}
\nabla^\mu\phi\nabla^\nu\phi-\frac{1}{2}\frac{\omega}{\phi^2}g^{\mu\nu}
\nabla^\alpha\phi\nabla_\alpha\phi
\end{equation}
It can be easly seen that the equation of the conformal factor is the
correct one at the first approximation. The quantum mass which appear in the
right hand side of equation (\ref{a7}) is consists of two parts. One
(the first term) has purely quantum nature and the other (the other terms)
has a mixture of quantum and cosmological aspect. This confirms the
 Machian's view point that we accepted here firstly. The continuty
equation is unchanged and the modified Eientien equations are the
same as the Brans-Dicke theory.
It must be noted that in refrence\cite{STQ} the dependence of quantum potential
and ensemble density is preassumed as before. In our present work that is
explained in the next section we want to remove this preassumption.

Also in another work\cite{QG}, a fully metric theory is constructed with results highly similar to
those of this scalar--tensor theory.
The field equations in that theory are in fact highly nonlinear and higher derivatives of the space--time
metric are included. So the scalar--tensor theory is replaced with a tensor theory at the cost of complexifying the field equations.
The positive point of this work is that the quantum effects of vacuum, (i.e. when no matter is present), can be calculated.
This is done for a black--hole in \cite{QG}.
\section{QUANTUM GRAVITY AND NON--MINIMAL SCALAR--TENSOR THEORIES}
In the present work, we want to write an appropriate action such that the conformal factor and quantum potential can be assumed as dynamical fields. In this way the relation between the conformal factor
and quantum potential and also the dependence of quantum potential to the ensemble density are
resulted at the first order of approximation. The arguments of the previous section leads us to use
a non--minimal scalar--tensor action. Thus we start from the most general non--minimal action:
\begin{equation}
{\cal A}=\int \! d^4x\ \sqrt{-g}\left [ \phi{\cal R}-\omega\frac{\nabla_\mu\phi\nabla^\mu\phi}{\phi} -\frac{\nabla_\mu Q\nabla^\mu Q}{\phi}+2\Lambda\phi+{\cal L}_m\right ]
\end{equation}
where $\Lambda$ is the cosmological constant, that generally have an interaction term with the
scalar filed.
We prefer to use the matter lagrangian:
\begin{equation}
{\cal L}_m=\frac{\rho}{m}\phi^a\nabla_\mu S\nabla^\mu S-m\rho\phi^b-\Lambda(1+Q)^c
+\alpha\rho(e^{\beta Q}-1)
\end{equation}
The first three terms of this lagrangian are the same as those of our previous work\cite{STQ}. The last
term is chosen in such a way, that satisfies two facts. It is necessary to have an interaction between the
quantum potential field and the ensemble density, to have a relation between them via the equations of motion.
Furthuremore, this interaction is written such that in the classical limit, it vanishes.

Variation of the above action functional leads to the following equations of motion:
\begin{itemize}
\item {\bf the scalar field's equation of motion}
\begin{equation}
{\cal R}+\frac{2\omega}{\phi}\Box\phi-\frac{\omega}{\phi^2}\nabla^\mu\phi\nabla_\mu\phi
+2\Lambda+\frac{1}{\phi^2}\nabla^\mu Q\nabla_\mu Q+\frac{a}{m}\rho\phi^{a-1}\nabla^\mu S
\nabla_\mu S-mb\rho\phi^{b-1}=0
\label{a8}
\end{equation}
\item {\bf the quantum potential's equation of motion}
\begin{equation}
\frac{\Box Q}{\phi}-\frac{\nabla_\mu Q\nabla^\mu\phi}{\phi^2}-\Lambda c(1+Q)^{c-1}
+\alpha\beta\rho e^{\beta Q}=0
\label{a11}
\end{equation}
\item {\bf the generalized Einstein's equation}
\[ {\cal G}^{\mu\nu}-\Lambda g^{\mu\nu}=-\frac{1}{\phi}{\cal T}^{\mu\nu}-\frac{1}{\phi}
\left [ \nabla^\mu\nabla^\nu-g^{\mu\nu}\Box\right ]\phi+\frac{\omega}{\phi^2}
\nabla^\mu\phi\nabla^\nu\phi-\frac{\omega}{2\phi^2}g^{\mu\nu}\nabla^\alpha\phi
\nabla_\alpha\phi \]
\begin{equation}
+\frac{1}{\phi^2}\nabla^\mu Q\nabla^\nu Q-\frac{1}{2\phi^2}
g^{\mu\nu}\nabla^\alpha Q\nabla_\alpha Q
\label{a9}
\end{equation}
\item {\bf the continuity equation}
\begin{equation}
\nabla_\mu\left (\rho\phi^a\nabla^\mu S\right )=0
\end{equation}
\item {\bf the quantum Hamilton--Jacobi equation}
\begin{equation}
\nabla^\mu S\nabla_\mu S=m^2\phi^{b-a}-\alpha m\phi^{-a} (e^{\beta Q}-1)
\label{a10}
\end{equation}
\end{itemize}
In equation (\ref{a8}), the scalar curvature and the term $\nabla^\mu S\nabla_\mu S$ can be eliminated
using the equations (\ref{a9}) and (\ref{a10}). In addition, on using the matter lagrangian and the definition of the energy--momentum tensor, one has:
\begin{equation}
(2\omega-3)\Box\phi=(a+1)\rho\alpha(e^{\beta Q}-1)-2\Lambda(1+Q)^c+2\Lambda\phi-\frac{2}{\phi}
\nabla_\mu Q\nabla^\mu Q
\label{a12}
\end{equation}
where the constant $b$ is chosen as $a+1$ as in the previous work\cite{STQ}.
We solve the equations (\ref{a11}) and (\ref{a12}), using perturbative expansion with $\alpha$ as the expansion
parameter:
\begin{equation}
Q=Q_0+\alpha Q_1+\cdots
\end{equation}
\begin{equation}
\phi=1+\alpha Q_1+\cdots
\end{equation}
\begin{equation}
\sqrt{\rho}=\sqrt{\rho}_0+\alpha\sqrt{\rho}_1+\cdots
\end{equation}
where the conformal factor is chosen to be unity at the zeroth order of perturbation, because in the limit
$\alpha\rightarrow 0$ the equation (\ref{a10}) would be leads the classical Hamilton--Jacobi equation.
Since by equation (\ref{a10}), the quantum mass is given by $m^2\phi+{\rm other\ terms}$,
the first order term of $\phi$ is chosen to be $Q_1$ as it must be so according to the relation of quantum mass (\ref{a6}). Also we shall show that $Q_1$ would be equal to $\Box\sqrt{\rho}/\sqrt{\rho}$ plus some corrections, which is desired as we called $Q$ the quantum potential field.

At the zeroth order one gets:
\begin{equation}
\Box Q_0-\Lambda c -\Lambda c(c-1)Q_0=0
\label{a13}
\end{equation}
\begin{equation}
\nabla_\mu Q_0\nabla^\mu Q_0=-\Lambda cQ_0
\label{a14}
\end{equation}
and at the first order:
\begin{equation}
\nabla_\mu Q_0\nabla^\mu Q_1=\Box Q_1-Q_1\Box Q_0-\Lambda c(c-1)Q_1+\beta\rho_0 e^{\beta Q_0}
\label{aaa}
\end{equation}
\begin{equation}
(2\omega-3)\Box Q_1=(a+1)\rho_0(e^{\beta Q_0}-1)-2\Lambda (c-1)Q_1-4\nabla_\mu Q_0\nabla^\mu Q_1
+2Q_1\nabla_\mu Q_0\nabla^\mu Q_0
\label{a15}
\end{equation}
On using equations (\ref{a13}),  (\ref{a14}) and (\ref{aaa}), in the equation (\ref{a15}), one gets:
\[ -(1+2\omega)\Box Q_1+2\Lambda Q_1\left ( (1-c+2c^2)+2c(c-3/2)Q_0\right ) \]
\begin{equation}
+[(a+1)(e^{\beta Q_0}-1)-4\beta e^{\beta Q_0}]\rho_0=0
\end{equation}
This equation can be written in the simple form:
\begin{equation}
\Box Q_1+A(\rho_0)Q_1+B(\rho_0)=0
\label{a16}
\end{equation}
where
\begin{equation}
A(\rho_0)=\frac{-1}{1+2\omega}2\Lambda \left ( (1-c+2c^2)+2c(c-3/2)Q_0\right )
\label{a17}
\end{equation}
\begin{equation}
B(\rho_0)=\frac{-1}{1+2\omega}[(a+1)(e^{\beta Q_0}-1)-4\beta e^{\beta Q_0}]\rho_0
\label{a18}
\end{equation}
The equation (\ref{a16}) can be solved iteratively. At the first iteration:
\begin{equation}
Q_1^{(1)}=-\frac{B}{A}
\end{equation}
and at the second and third iteration:
\begin{equation}
Q_1^{(2)}=\frac{1}{A}\Box\left ( \frac{B}{A}\right )-\frac{B}{A}
\end{equation}
\begin{equation}
Q_1^{(3)}=-\frac{1}{A}\Box\left ( \frac{\Box B/A}{A}\right ) +\frac{1}{A}\Box\left ( \frac{B}{A}\right )-\frac{B}{A}
\end{equation}
In order to have the correct dependence of the quantum potential on the ensemble density, it
is sufficient to set:
\begin{equation}
A=k_1\sqrt{\rho_0};\ \ \ \ \ \ \ \ \ \ B=k_2\rho_0
\label{a19}
\end{equation}
where $k_1$ and $k_2$ are two constants.
This leads to the following expressions for the quantum potential up to the third order of iteration:
\begin{equation}
Q_1^{(1)}=-\frac{k_2}{k_1}\sqrt{\rho_0}
\end{equation}
\begin{equation}
Q_1^{(2)}=\frac{k_2}{k_1^2}\frac{\Box\sqrt{\rho_0}}{\sqrt{\rho_0}}-\frac{k_2}{k_1}\sqrt{\rho_0}
\end{equation}
\begin{equation}
Q_1^{(3)}=-\frac{k_2}{k_1^3}\frac{1}{\sqrt{\rho_0}}\Box\left ( \frac{\Box\sqrt{\rho_0}}{\sqrt{\rho_0}}\right )
+\frac{k_2}{k_1^2}\frac{\Box\sqrt{\rho_0}}{\sqrt{\rho_0}}-\frac{k_2}{k_1}\sqrt{\rho_0}
\end{equation}
If the ensemble density be not much great, and it be so smooth that its higher derivatives be small,
the result would be in agreement with the desired relation $Q=\Box\sqrt{\rho_0}/\sqrt{\rho_0}$
provided we choose $k_2=k_1^2=k$.
Comparison of relations (\ref{a17}), (\ref{a18}) and (\ref{a19}) leads to:
\begin{equation}
a=2\omega k; \ \ \ \ \ \ \beta=\frac{2\omega k+1}{4};\ \ \ \ \ Q_0=\frac{1}{c(2c-3)}\left [ -\frac{2\omega k+1}{2\Lambda}k \sqrt{\rho_0}-(2c^2-c+1)\right ]
\label{a20}
\end{equation}
The space--time dependence of $\rho_0$ can be derived from the relation (\ref{a14}).

We see that the except $c$ and $\omega$, all other constants are fixed.  The other equations of motion which
are not used in the perturbation procedure can be used to determine the space--time metric and
the Hamilton--Jacobi function.

We conclude this section with emphasizing on the fact that in our present work, the quantum
potential is a dynamical field. And, that solving perturbatively the equations of motion one gets
the correct dependence of quantum potential upon density plus some corrective terms.
\section{CONCLUSION}
In this paper we first reviewed the recent works showing that quantum effects are nothing  else
the conformal degree of freedom of the space--time. Then we have construct a scalar--tensor
theory with two scalar fields, for which the equations of motion leads to the correct form of quantum potential.
In fact we have shown that using the perturbative solutions of the equations of motion and
from the Hamilton--Jacobi equation the quantum mass is given by:
\begin{equation}
{\cal M}^2=m^2\phi-\alpha m \phi^{-2\omega}(e^{(2\omega k+1)Q/4}-1)
\end{equation}
which up to first order in $\alpha$ is:
\begin{equation}
{\cal M}^2=m^2(1+\alpha\Box\sqrt{\rho_0}/\sqrt{\rho_0})-
\frac{\alpha m^2}{k}\sqrt{\rho_0}-\alpha m(e^{(2\omega k+1)Q_0/4}-1)
\end{equation}
in which $Q_0$ is given by the relation (\ref{a20}).
As we see the mass function includes the correct term (\ref{a21}) and some corrective terms.
This seems to be a great succes. Since an essential question in Bohm's theory is that why the quantum potential
has that strange form. In this work, we have not only show that quantum effects are gemetrical
in nature, but also {\it derive\/} the form of quantum potential. This specific form for quantum
potential is a result of the equations of motion.

It must be noted that in this theory both the scalar fields interacts with the cosmological constant.
So the presence of the cosmological constant (even very small) is essential in order the theory works.
Note that the interaction between $\Lambda$ and $Q$ represents a connection between the large
scale ($\Lambda$) and the small scale ($Q$) structures.

\end{document}